\documentclass[letterpaper, 10 pt, conference]{ieeeconf}  %
\IEEEoverridecommandlockouts      
\usepackage{cite}
\usepackage{amsmath,amssymb,amsfonts}
\usepackage{algorithmic}
\usepackage{graphicx}
\usepackage{textcomp}
\usepackage{xcolor}

\usepackage{booktabs}
\usepackage{multirow}
\usepackage{xspace}
\usepackage{color}
\usepackage[normalem]{ulem}
\useunder{\uline}{\ul}{}

\usepackage[T1]{fontenc}
\usepackage{aecompl}

\usepackage{caption}
\usepackage{subcaption}

\let\labelindent\relax
\usepackage{enumitem}

\newcommand{\ours}{STATENet\xspace}

\def\x{{\mathbf x}}
\def\h{{\mathbf h}}
\def\w{{\mathbf w}}
\def\m{{\mathbf m}}
\def\L{{\cal L}}

\def\BibTeX{{\rm B\kern-.05em{\sc i\kern-.025em b}\kern-.08em
    T\kern-.1667em\lower.7ex\hbox{E}\kern-.125emX}}

\title{Protecting the Future: Neonatal Seizure Detection\\with Spatial-Temporal Modeling
}

\author{Ziyue Li$^{1,*}$,
    Yuchen Fang$^{1,2}$, 
    You Li$^{1,3}$, 
    Kan Ren$^{1,\dagger}$, 
    Yansen Wang$^1$, 
    Xufang Luo$^1$,\\
    Juanyong Duan$^1$, 
    Congrui Huang$^1$, 
    Dongsheng Li$^1$, 
    Lili Qiu$^1$\\
    \\
    https://seqml.github.io/statenet/
\thanks{$^{1}$Microsoft}%
\thanks{$^{2}$Shanghai Jiao Tong University}%
\thanks{$^{3}$Central South University}%
\thanks{$^*$
    The first two authors share co-first authorship. The work was done during the internship of Ziyue Li, Yuchen Fang, and You Li at Microsoft Research Asia.
    }
\thanks{$^\dagger$
    Correspondence to Kan Ren <kan.ren@microsoft.com>.
    }
}

\begin{document}

\maketitle
\thispagestyle{empty}
\pagestyle{empty}

\begin{abstract}
A timely detection of seizures for newborn infants with electroencephalogram (EEG) has been a common yet life-saving practice in the Neonatal Intensive Care Unit (NICU). 
However, it requires great human efforts for real-time monitoring, which calls for automated solutions to neonatal seizure detection. 
Moreover, the current automated methods focusing on adult epilepsy monitoring often fail due to (i) dynamic seizure onset location in human brains;
(ii) different montages on neonates and (iii) huge distribution shift among different subjects.
In this paper, we propose a deep learning framework, namely
\ours, to address the exclusive challenges with exquisite designs at the temporal, spatial and model levels. 
The experiments over the real-world large-scale neonatal EEG dataset illustrate that our framework achieves significantly better seizure detection performance.
\end{abstract}

\section{Introduction}
\label{sec:intro}
As neurological disorders caused by epilepsy, seizures are associated with high morbidity and mortality for new born infants \cite{rakshasbhuvankar2015amplitude}. And the timely detection and proper treatments become a common yet vital practice in the Neonatal Intensive Care Unit (NICU). Although it can be observed and detected through electroencephalogram (EEG) of an individual which has been considered as the ``golden standard", it requires significant human efforts from experts for monitoring, caring and intensive diagnosis. 
Thus, building an accurate automated framework to detect seizure events in real time can help free experts from tedious works to better focus on the treatments.

Automated seizure detection attracts lots of attention in both signal processing and machine learning (ML) communities. 
Signal processing methods \cite{khamis2013frequency,bigdely2015prep,jas2017autoreject} mainly focus on statistical or hand-crafted features of EEG, requiring lots of expert knowledge. 
Besides, ML approaches \cite{tang2021self,zheng2021predicting,9053143} rely on data-driven paradigm to process EEG for seizure detection.
But these solutions are mainly focusing on adult epilepsy monitoring and can not be directly applied to neonates considering several clinical differences. 
While some deep learning methods \cite{gramacki2022deep,8168193,o2021deep} were proposed for EEG seizure detection in neonates, they did not explicitly handle the critical challenge in this problem, i,e, dynamics in the number of electrodes and seizure patterns, which often introduces more artifacts as burdens of detecting informative seizure events.

To address these exclusive challenges, we propose a deep learning framework, SpaTiAl-Temporal EEG Network (\ours). In \ours, we incorporate a channel-level temporal modeling component for fine-grained brain signal processing, which is more flexible when tackling varying yet limited EEG channels on neonates. After the temporal modeling process, we leverage a spatial fusion module to comprehensively synthesize channel-level temporal patterns for detection.
This process has been optimized through an end-to-end manner without explicitly signal preprocessing or human-crafted artifact removal.
Moreover, we propose a model-level ensemble by dynamically aggregating the outcomes of diverse spatial-temporal deep models to better generalize among different neonates.
We conduct experiments on a real-world large-scale neonatal dataset, to compare our model with several competitive baselines focusing on adult seizure detection and illustrate that our method has achieved significantly better seizure detection performance. 
Furthermore, we limit the number of channels to simulate a different clinical scenario and transfer the learned model directly, and we observe little performance drop in the detection quality of \ours, which illustrates the robustness of our method.

\section{Preliminaries}
\label{sec:prem}

\begin{figure*}
\centering
\begin{subfigure}{.49\linewidth}
  \centering
  \includegraphics[width=0.7\textwidth]{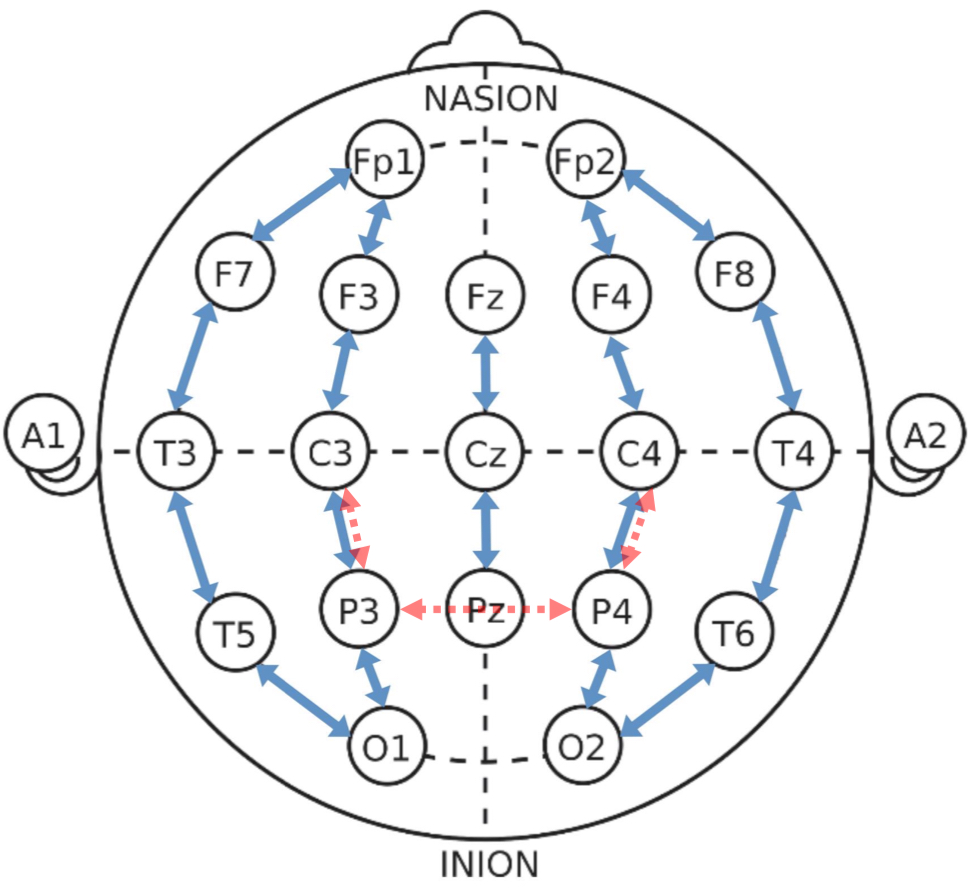}
  \caption{
      The two bipolar montages of electrode graph. The blue arrows represent 18-channel one and the red arrows are 3-channel montages.
      }
  \label{fig:elec-graph}
\end{subfigure}
\hfill
\begin{subfigure}{.49\linewidth}
  \centering
  \includegraphics[width=0.7\textwidth]{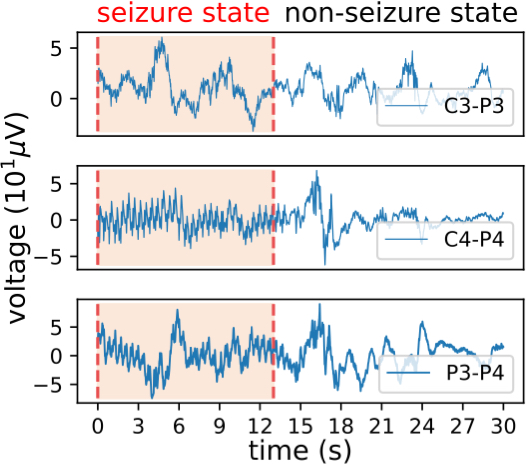}
  \caption{EEG signal sample visualization with epileptic seizure state in highlighted time range annotated by human experts. 
  }
  \label{fig:eeg-wave}
\end{subfigure}
\caption{Visualization of the bipolar montages and the EEG signal samples with seizure onset.}
\end{figure*}

\subsection{Materials}
\label{sec:materials}
In this paper, we utilize the neonatal EEG recording dataset \cite{stevenson2019dataset}, which collected multi-channel EEG signals from a cohort of 79 term neonates admitted to the NICU at the Helsinki University Hospital.
The recordings have been annotated by three experts individually and an average of 460 seizures were annotated per expert in the dataset; 
Among them, 39 neonates had seizures and 22 were seizure free, by consensus of all experts.

The dataset provides a standard 18-channel bipolar montage \cite{stevenson2019dataset}, with the electrode graph of solid blue arrows illustrated in Fig.~\ref{fig:elec-graph}.
Note that, the head size of neonates is relatively smaller, which may make the full montage, that is successfully tested on adults, inappropriate for neonates.
Thus, additionally, we select 3-channel bipolar montage: C3-P3, C4-P4 and P3-P4, following the findings of human experts \cite{tao2010using,lawrence2009pilot}, as illustrated in red dotted line annotation of Fig.~\ref{fig:elec-graph}.
As an example shown in Fig.~\ref{fig:eeg-wave}, the pieced EEG waves of the selected three bipolar channels contain normal state and epileptic waving state. 
In seizure state (with orange area), the brain signal waves of all the channels present apparent disorder; while in the non-seizure state, the signal waves recover to normal fluctuation.

\subsection{Solution Formulation}
The whole dataset consists of $N$ EEG samples $\{ (\x^{(i)}, y^{(i)}) \}_{i=1}^N$ with seizure labels.
Without cause of confusion, we omit the index notation $i$ for clearer clarification.

The goal of seizure detection is to estimate the probability $\hat{p}=\text{Pr}(y=1 |\x )$ that there exists a seizure state ($y=1$) in the current time piece given the EEG wave sample $\x$, where $y\in\{0,1\}$.
The input 
$\x=[\x_1, \ldots, \x_c, \ldots, \x_C]^\top$ 
is a multivariate time-series instance containing totally $C$ channels of EEG time-series signals, as illustrated in Fig.~\ref{fig:eeg-wave}.
Each univariate time series is $\x_c \in \mathbb{R}^L$ where $L$ is the overall timestep number.
Concretely, $L=6000$ for the $30$-second EEG signal recorded in $200$Hz sampling frequency.

Formally, given the sample input $\x$, each model $f(\cdot; \theta)$ with parameter $\theta$ estimates the seizure probability as $\hat{p}=f_\theta(\x)$.
The learning objective is to minimize the cross-entropy loss w.r.t. the model parameter $\theta$ as
\begin{equation}\label{eq:loss}
\begin{split}
\L = - \frac{1}{N}\sum_{i=1}^N \left[ y^{(i)}\log\hat{p}^{(i)} + (1-y^{(i)}) \right. & \left. \log(1-\hat{p}^{(i)}) \right] \\
            &+ \lambda \frac{1}{2}\|\theta\|^2 ,
\end{split}
\end{equation}
with the regularization term weighted by hyperparameter $\lambda$.

\subsection{Challenges of Neonatal Seizure Detection}
\label{sec:challanges}
Here we describe the existed challenges in processing and modeling the EEG signal data for neonatal seizure detection, which also motivates our model design detailed later.

\textbf{Challenge 1: Epilepsy seizure events occur dynamically in different channels.}
In Fig.~\ref{fig:eeg-wave}, when in the seizure state, the EEG waves illustrate typical disorder with high-frequency spike and wave discharges.
However, the signals $\x_c$ at different channels have shown different patterns or even do not illustrate seizure, which corresponds to different causes of epileptic seizure and their paroxysm location in the brain.
The existing methods such as \cite{tang2021self} often overlook the fine-grained seizure pattern situation and takes the signals of all channels as a whole which may confound various patterns thus degenerate the detection performance.
Thus, fine-grained EEG signal processing is required.

\textbf{Challenge 2: The channel number is variant even limited on neonatal brain health monitoring.}
Since the head size of neonates is much smaller than adults, which results in the limited sensing electrodes limited in the real scenario.
And the sensing devices in different centers can be quite different, which leads to variant channel information in the data.
Several related works \cite{lawrence2009pilot,rakshasbhuvankar2015amplitude,variane2021current} have also studied neonatal seizure detection with limited channels, e.g., only two channels.
All the observations encourage researchers to conduct dynamic modeling techniques for dynamically modeling various even limited EEG signal channels for seizure detection.

\begin{figure}
    \centering
    \includegraphics[width=1.\columnwidth]{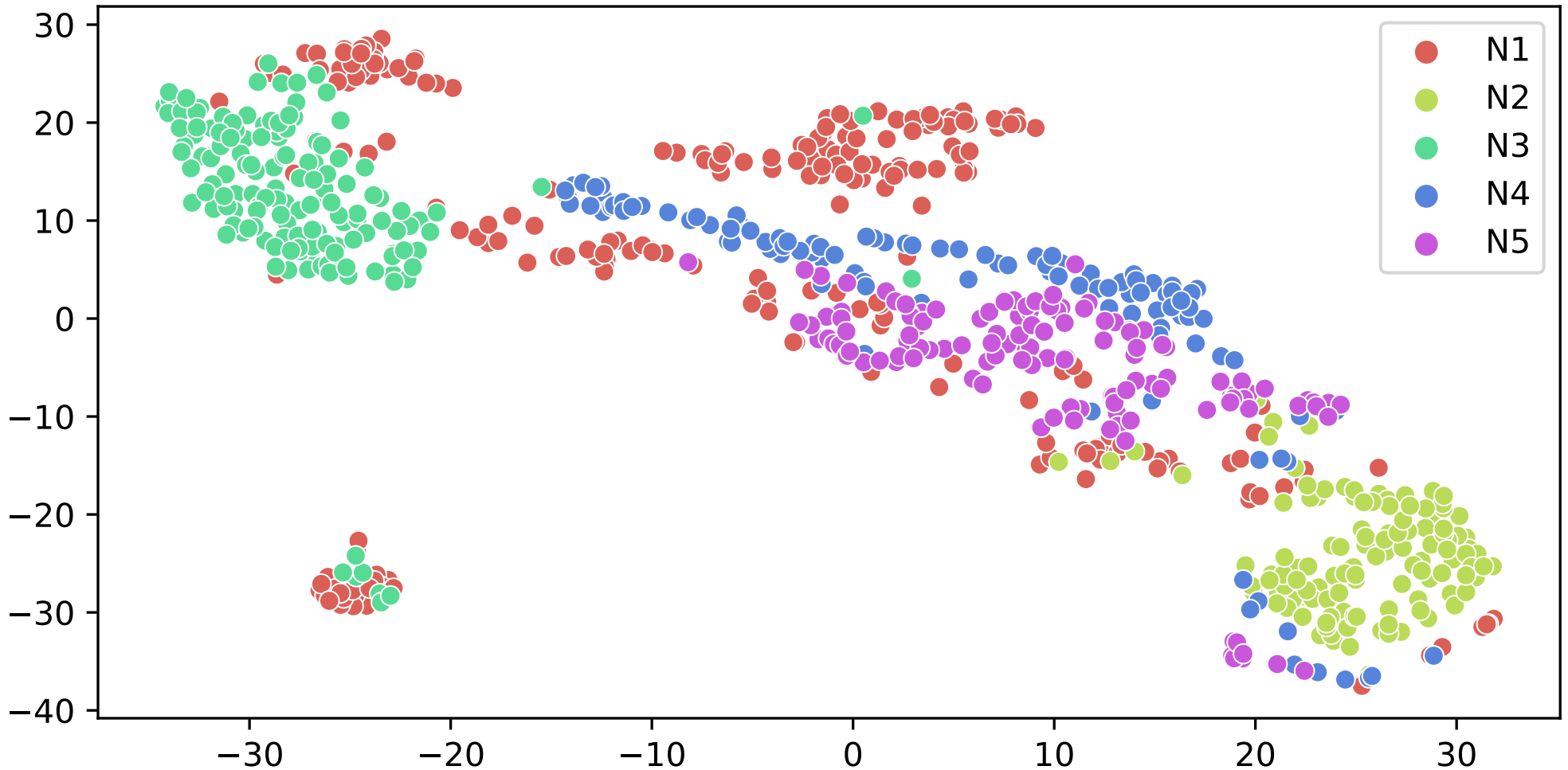}
    \caption{The t-SNE \cite{van2008visualizing} visualization of 5 different neonates.}
    \label{fig:cluster}
\end{figure}

\textbf{Challenge 3: Seizure patterns vary among neonates resulting in model generalization issue.}
The EEG dataset often contains the recordings from a cohort of people and a large variance of data distributions from different subjects have been observed, as shown in Fig.~\ref{fig:cluster}.
From the figure, we can find that different clusters representing the EEG signal distributions of different neonates diverge in a large margin, which places obstacles for machine learning models to generalize from the training dataset to the test dataset which violates the independently identical distributional assumption of machine learning.

\section{Method}
\label{sec:method}

\subsection{\ours: Spatial-Temporal EEG Network}
\label{sec:framework}
For each input EEG montage $\x$, we first conduct a channel-level temporal modeling that independently models the signal of each channel.
To get the final prediction $\hat{p}$ and dynamically detect seizure from different channels, we then utilize a multi-channel spatial fusion module to fuse the information from multiple channels, as described below.
This proposed framework manages to conduct fine-grained modeling on single channel while remains feasible to handle the input with various channel numbers, tackling above Challenges 1 and 2.

\noindent\textbf{Channel-level Temporal Modeling.}
\label{sec:temporal}
We utilize temporal convolutional network (TCN)~\cite{bai2018empirical} for channel-level temporal modeling.
By stacking $L$ dilated convolution layers with 1-$d$ convolution filters, we independently model each channel with relatively low time complexity.
The process can be formulated as
\begin{equation}\label{eq:temporal}
    \h_c^0 = \x_c, ~~
    \h_c^l = \text{ReLU}(h_c^{l-1} \ast \mathbf{w}_l),  ~~
    \h_c = \h_c^L~,
\end{equation}
where $\mathbf{w}_l$ is the convolution filter of the $l$-th layer, which is shared among all channels, $\ast$ is the dilated convolution operation, $\h_c^l$ is the extracted hidden representation of the $c$-th channel at the $l$-th layer, and $\text{ReLU}(x) = \text{max}(x, 0)$ is the activation function.

\noindent\textbf{Multi-channel Spatial Fusion.}
\label{sec:spatial}
We utilize a multi-channel spatial fusion module to fuse the information from each channel as
\begin{equation}\label{eq:spatial}
    \hat{p} = f_s(g([\h_1, \h_2, ..., \h_C]))~,
\end{equation}
where $g:\mathbb{R}^{C \times d} \mapsto \mathbb{R}^d$ is an aggregation function and $f_s$ is a three-layer multi-layer perception (MLP) with ReLU as activation function.

The EEG signal of each channel reflects the brain activity in a certain brain region.
It is important to model the spatial relations between these channels to better detect unusual EEG patterns and locate the source of seizure.
However, existing methods either overlook the spatial relations between channels or relies on a predefined graph to describe the connections between channels~\cite{tang2021self}, which is unavailable under neonatal EEG scenario when the number and positions of channels are not fixed. 
To solve this problem, we utilize Graph Neural Networks (GNN) as our aggregation function $g$, which are commonly used for spatial-temporal modeling~\cite{wu2020conditional, fang2023learning, li2017diffusion} including EEG modeling~\cite{tang2021self}.
Specifically, we implement the aggregation function $g$ as a graph attention network (GAT) \cite{velivckovic2017graph} to dynamically decide the strength of spatial connectivity between channels.
The process is denoted as
\begin{align}
    &\m_c^0 = \h_c, \\
    &\alpha_{i, j}^l = \frac{e^{\mathbf{W}\m_i^l \cdot \mathbf{W}\m_j^l}}{\sum_{k=1}^C e^{\mathbf{W}^l\m_i^l \mathbf{W}\cdot \m_k^l}}, 1 \leq i,j \leq C, \\ 
    &\m_c^{l+1} = \sum_{k=1}^C \alpha_{c,k}^l\mathbf{W}\m_k^l,\\
    &g[\h_1, \h_2, ..., \h_C] = \frac{1}{C} \sum_c \m_c^L,
\end{align}
where $0 \leq l \leq L$ is the layer of GAT, $\m_c^l$ is the node representation vector of the $c$-th channel at the $l$-th GAT layer, $\mathbf{W} \in \mathbb{R}^{d \times d}$ are learnable parameters, and $\alpha$ is the edge weights dynamically decide by the node representations of the last layer.
This module flexibly models the spatial relations between changing number of channels and efficiently aggregates the information from different channels.
The dynamic spatial relation mining of $g$ also helps locating EEG seizure source, as we will exhibit in Sec.~\ref{sec:ext-invest}.

\subsection{Mixture of Experts for Cross-Person Generalization}
\label{sec:ensemble}

To tackle the generalization issue of Challenge 3 described in Sec.~\ref{sec:challanges}, we incorporate diverse models which own different specialties and utilize a mixture-of-expert (MoE) framework ~\cite{jacobs1991adaptive} to specify their contributions to predicting each test sample.
Ensemble learning is known as an effective method to improve the generalization ability of neural networks~\cite{saikia2019investigating,sadat2021alzheimer}.
Meanwhile, MoE has demonstrated excellent performance in computer vision~\cite{lisimple}, natural language processing~\cite{shazeer2017outrageously}, etc.
In addition, although multiple models are included, the computational cost of ensemble is manageable and can even be compared to or lower than that of a single model~\cite{li2023towards}.

Specifically, each sample $\x^{(i)}$ is assigned to $K$ models, and their predictions $[\hat{p}_{1}^{(i)}, \dots, \hat{p}_{K}^{(i)}]$ are weighted by normalized \textit{sample-level ensemble weights} $\w^{(i)} \in \mathbb{R}^K$ to output the ensemble prediction, which is obtained by $ \hat{p}^{(i)} = \sum^{K}_{k=1} \w^{(i)}_{k} \hat{p}_{k}^{(i)}$ where $\sum^{K}_{k=1} \w^{(i)}_{k} = 1$. 

To obtain its ensemble weights, each sample $\x^{(i)}$ is first embedded using a standard GRU network~\cite{cho2014learning}, and the output is then fed into a single-layer MLP. The MLP output is normalized using the softmax operation $\text{Softmax}(\mathbf{z})_k={e^{z_k}}/{\sum_{j=1}^K e^{z_j}}$ to obtain the sample-level ensemble weights
\begin{equation}
\w^{(i)} = \text{Softmax}\left[\text{MLP}(\text{GRU}(\x^{(i)}))\right].
\end{equation}
Note that the integrated base models are not fine-tuned, and only the parameters of the GRU network and the single-layer MLP are updated.
In Sec.~\ref{sec:ext-invest}, we demonstrate that despite incurring additional computational costs, our proposed ensemble yields significant performance gains.

Our proposed ensemble method offers several advantages over existing methods.
Firstly, our approach leverages a diverse set of models, allowing for greater flexibility in capturing a wide range of seizure patterns. 
Secondly, our ensemble method dynamically dispatches models based on a given neonatal sample, enhancing the adaptability of ensembles to individual differences. 
Moreover, our approach does not require complex calibration or training phases, making it easy to implement in existing clinical settings.

\section{Experiment}\label{sec:exp}

\subsection{Experimental Setup}
\noindent\textbf{Dataset Preparation.}
The dataset \cite{stevenson2019dataset} presented in Sec.\ref{sec:materials} consists of the EEG records of 79 neonates, and we randomly split the dataset into training and test sets by patients following \cite{tang2021self}, which is realistic since the automated seizure detection service is trained on existing data and predicts for future coming patients.
Based on the data split, we conduct four-fold cross-validation for evaluation.
We further obtain 30-s EEG clips using non-overlapping sliding windows and overall sample number is more than 40 thousand.
And each clip is annotated as a positive sample if all the three experts annotated the presence of seizure.

\noindent\textbf{Compared Methods.}
We compare our method with existing seizure detection and time series classification methods, including 
\textit{GBDT}~\cite{prokhorenkova2018catboost}, a gradient boosting decision tree model utilizing spike, temporal feature extracted from EEG data;
\textit{ROCKET}~\cite{dempster2020rocket}, which utilizes random convolution kernels to extract feature vectors from EEG data and uses a ridge regression to get final predictions;
\textit{GRU}~\cite{cho2014learning}, a recurrent neural network (RNN) with a gating mechanism to efficiently capture information in long signals;
\textit{TCN}~\cite{bai2018empirical}, a dilated convolutional neural network (CNN) designed for time-series modeling;
\textit{MLSTM-FCN}~\cite{karim2019multivariate} combining CNN and RNN with squeeze-and-excitation blocks;
\textit{InceptionTime}~\cite{ismail2020inceptiontime} enhancing CNN and RNN with time-series Inception modules;
\textit{DCRNN}~\cite{tang2021self} utilizing correlation between variables to build a graph to capture relations between brain activities and leveraging Fourier transform to capture meaningful information in EEGs for seizure detection and classification.
All methods are evaluated using area under precision-recall curve (AUPRC) and area under receiver operating characteristic curve (AUROC).
For both metrics, higher value indicates better performance.
Each reported number is averaged from three runs of different random seeds.
We will publish the codes upon the acceptance of this paper.

\noindent \textbf{Setting of Ensemble.}
To enhance the generalization capability of our neural network models, we propose an ensemble method that leverages diverse network architectures. 
Specifically, our ensemble consists of four independently trained models, each based on a different network architecture: GRU, DCRNN, TCN, and \ours. 
These models are dispatched to different test samples by the MoE framework.

\subsection{Experimental Results}

\begin{table*}[ht]
\centering
\caption{The detailed experiment results. The higher metric value is better. The best and the second-placed results are formatted as bold font and underlined format, respectively.}
\begin{tabular}{ccrrrrrrrr|rr}
\toprule
\multirow{2}{*}{Method} & \multirow{2}{*}{Metric} & \multicolumn{2}{c}{fold1} & \multicolumn{2}{c}{fold2} & \multicolumn{2}{c}{fold3} & \multicolumn{2}{c|}{fold4} & \multicolumn{2}{c}{average} \\ \cmidrule(l){3-12} 
 &  & \multicolumn{1}{c}{3} & \multicolumn{1}{c}{18} & \multicolumn{1}{c}{3} & \multicolumn{1}{c}{18} & \multicolumn{1}{c}{3} & \multicolumn{1}{c}{18} & \multicolumn{1}{c}{3} & \multicolumn{1}{c|}{18} & \multicolumn{1}{c}{3} & \multicolumn{1}{c}{18} \\ \midrule
\multirow{2}{*}{GBDT} & AUROC & 72.1 & 73.8 & 76.6 & 76.4 & 81.3 & 78.8 & 81.0 & 79.7 & 77.7 & 77.2 \\
 & AUPRC & 37.4 & 36.2 & 45.5 & 46.2 & 56.5 & 54.5 & 57.9 & 47.4 & 49.4 & 46.1 \\ \midrule
\multirow{2}{*}{ROCKET} & AUROC & 75.7 & 64.8 & 76.4 & 73.1 & 86.0 & 80.3 & {\ul 85.4} & {\ul 84.4} & 80.9 & 75.7 \\
 & AUPRC & 49.2 & 37.7 & 43.1 & 32.4 & 60.5 & 59.4 & 59.9 & 50.4 & 53.2 & 45.0 \\ \midrule
\multirow{2}{*}{GRU} & AUROC & 61.9 & {\ul 82.5} & 51.7 & 74.7 & 58.7 & 80.1 & 81.0 & 78.8 & 63.3 & 79.0 \\
 & AUPRC & 30.2 & {\ul 59.5} & 13.4 & 51.9 & 27.4 & 65.2 & 55.0 & 45.4 & 31.5 & 55.5 \\ \midrule
\multirow{2}{*}{TCN} & AUROC & 77.7 & 79.3 & 75.8 & 75.6 & \textbf{90.2} & {\ul 89.0} & 85.3 & 82.1 & {\ul 82.3} & {\ul 81.5} \\
 & AUPRC & {\ul 54.1} & 54.5 & 52.9 & {\ul 54.1} & {\ul 76.2} & {\ul 76.1} & {\ul 63.3} & 53.1 & {\ul 61.6} & {\ul 59.5} \\ \midrule
\multirow{2}{*}{MLSTM-FCN} & AUROC & {\ul 78.3} & 77.2 & 75.6 & 74.2 & 85.5 & 85.9 & 84.0 & 81.2 & 80.9 & 79.6 \\
 & AUPRC & 51.2 & 48.2 & 48.7 & 49.0 & 61.4 & 71.1 & 61.4 & 51.6 & 55.7 & 55.0 \\ \midrule
\multirow{2}{*}{InceptionTime} & AUROC & 72.6 & 74.0 & 75.2 & 73.1 & 72.6 & 85.6 & 70.9 & 81.2 & 72.8 & 78.5 \\
 & AUPRC & 33.7 & 37.3 & 49.7 & 49.9 & 47.6 & 68.9 & 30.7 & 51.8 & 40.4 & 52.0 \\ \midrule
\multirow{2}{*}{DCRNN} & AUROC & 73.3 & 79.5 & {\ul 80.7} & {\ul 76.9} & 84.5 & 87.0 & 85.3 & 81.9 & 81.0 & 81.3 \\
 & AUPRC & 42.8 & 50.8 & {\ul 59.0} & 51.6 & 64.5 & 66.6 & 59.2 & {\ul 57.5} & 56.4 & 56.6 \\ \midrule
\multirow{2}{*}{\ours} & AUROC & \textbf{87.1} & \textbf{91.5} & \textbf{85.2} & \textbf{87.7} & {\ul 89.5} & \textbf{93.4} & \textbf{88.2} & \textbf{91.2} & \textbf{87.5} & \textbf{91.0} \\
 & AUPRC & \textbf{70.1} & \textbf{78.0} & \textbf{71.5} & \textbf{74.8} & \textbf{78.8} & \textbf{85.3} & \textbf{74.1} & \textbf{77.4} & \textbf{73.6} & \textbf{78.9} \\ 
 \midrule
\specialrule{0em}{1pt}{1pt} 
\hline
\multicolumn{1}{l}{\multirow{2}{*}{MoE Ensemble}} 
& \multicolumn{1}{l}{AUROC} & \multicolumn{1}{l}{89.4} & \multicolumn{1}{l}{92.7} & \multicolumn{1}{l}{88.6} & \multicolumn{1}{l}{87.3} & \multicolumn{1}{l}{91.1} & \multicolumn{1}{l}{96.1} & \multicolumn{1}{l}{88.9} & \multicolumn{1}{l|}{91.3}  & \multicolumn{1}{l}{89.5} & \multicolumn{1}{l}{91.8} \\
\multicolumn{1}{l}{} 
& \multicolumn{1}{l}{AUPRC} & \multicolumn{1}{l}{74.1} & \multicolumn{1}{l}{81.3} & \multicolumn{1}{l}{72.4} & \multicolumn{1}{l}{73.3} & \multicolumn{1}{l}{81.4} & \multicolumn{1}{l}{89.0} & \multicolumn{1}{l}{73.7} & \multicolumn{1}{l|}{77.6} & \multicolumn{1}{l}{75.4} & \multicolumn{1}{l}{80.3} \\ \bottomrule
\end{tabular}

\label{tab:result}
\end{table*}

The experiment results of all cross-validation folds and the average performance on both 18-channel and 3-channel datasets are presented in Table~\ref{tab:result}.
We have the following observations from the results.
(1) \textit{Superiority of individual model}: 
Our proposed \ours achieves the best performance compared with other baseline models without ensemble on most folds of two datasets, which illustrates that the fine-grained channel-level temporal modeling and spatial fusion offer great capacity for EEG modeling.
(2) \textit{Advances of ensemble}:
Over a diverse set of trained models, the ensemble model further boosts the performances, 
which results from the better generalization ability brought by the ensemble learning process.
(3) \textit{Transferability}: 
\ours achieves a comparable 
performance on 3-channel datasets 
to that on 18-channel datasets, indicating that \ours manage to adapt to limited channel scenario, thus more suitable for neonatal seizure detection. This observation is also consistent with clinical observations~\cite{variane2021current}.

\subsection{Extended Investigation}\label{sec:ext-invest}
\subsubsection{Transfer Across Montages}
Note that in Eq.~\eqref{eq:temporal}, the filters of our dilated convolution layers are shared across channels, and the spatial fusion operation in  Eq.~\eqref{eq:spatial} can adapt to montages with variant number of channels.
As a result, our \ours method can be easily transferred to EEG data with different channels \textit{without} retraining.
Table~\ref{tab:transfer} presents the results of the transferred models.
The results show that although the transferred models suffer performance drop compared with the results in Table~\ref{tab:result}, the performances are still comparable and outperform all our baselines.
It indicates that our spatial-temporal modeling framework offers great flexibility and is perfectly suitable for the neonatal EEG scenarios where the channel number is variant or even limited.

\begin{table*}[]
\centering
\caption{The results of transferred models. ``3 to 18'' means directly transferring the model trained on 3-channel data to 18-channel data, similarly for ``18 to 3''.}
\begin{tabular}{@{}cccccccccc|cc@{}}
\toprule
\multirow{2}{*}{Method} & \multirow{2}{*}{Metric} & \multicolumn{2}{c}{fold1} & \multicolumn{2}{c}{fold2} & \multicolumn{2}{c}{fold3} & \multicolumn{2}{c|}{fold4} & \multicolumn{2}{c}{average} \\ \cmidrule(l){3-12} 
 &  & 18 to 3 & 3 to 18 & 18 to 3 & 3 to 18 & 18 to 3 & 3 to 18 & 18 to 3 & 3 to 18 & 18 to 3 & 3 to 18 \\ \midrule
\multirow{2}{*}{\ours} & AUROC & 82.3 & 90.1 & 83.0 & 89.1 & 80.0 & 91.9 & 85.7 & 86.9 & 82.8 & 89.5 \\
 & AUPRC & 58.3 & 75.5 & 66.3 & 75.7 & 68.4 & 76.8 & 71.4 & 65.4 & 66.1 & 73.4 \\ \bottomrule
\end{tabular}
\label{tab:transfer}
\end{table*}

\subsubsection{Occlusion Map Based Localization}
\begin{figure*}
    \centering
    \includegraphics[width=1.25\columnwidth]{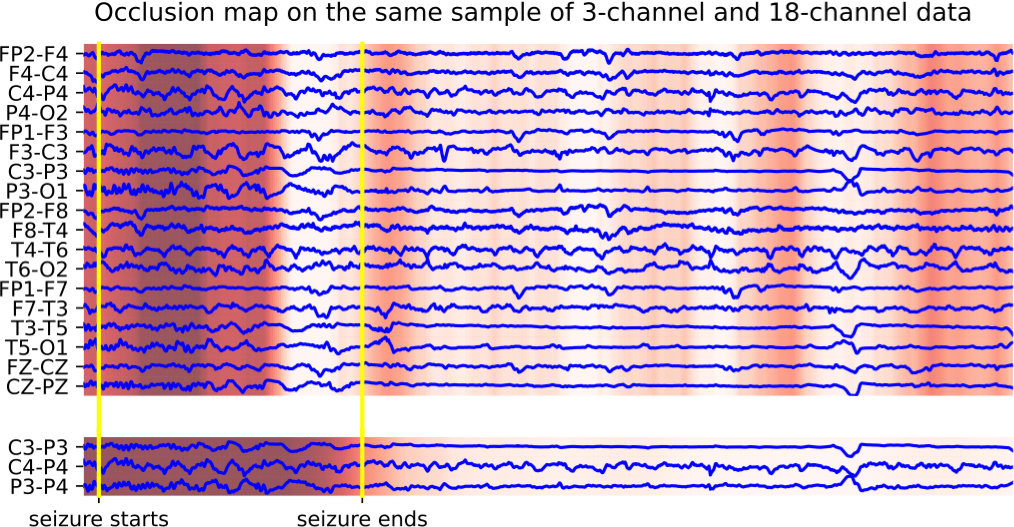}
    \caption{The occlusion map of \ours model on the same sample on 3-channel and 18-channel datasets. The deeper the color denotes more contribution to the positive prediction. The annotations of experts are shown with yellow lines.}
    \label{fig:occlusion}
\end{figure*}

We leverage occlusion map techniques to analyze the localization ability of our models.
Figure~\ref{fig:occlusion} shows the occlusion map of a same sample from 3-channel and 18-channel datasets, using the prediction of \ours trained on 18-channel dataset.
We observe that both occlusion maps indicate that the seizure occurs at the early phase of this sample, which accord with the annotations of experts.
The occlusion maps show that our \ours method has good interpretability and can help for seizure localization, and the ability is also transferable across data with varying number of channels.

\subsubsection{Model Suitability}
\begin{figure*}
    \centering
    \includegraphics[width=1\columnwidth]{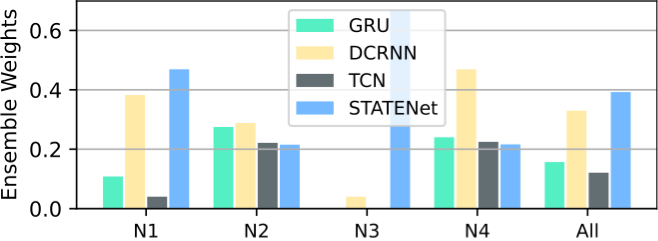}
    \caption{Average ensemble weights assigned to models 
    for four neonates (N1 to N4) and all neonates (All).}
    \label{fig:ens_weight}
\end{figure*}

Motivated by the evidence that data distributions differ significantly from people, we design a mixture-of-expert based framework that dynamically assigns models to each sample.
Here we analyze its dispatching behaviors to reveal whether models are diversified to different distributions in terms of people.
Specifically, we randomly select 4 neonates, on which we compute ensemble weights assigned to each model averagely by our best-performed ensemble, with results shown in Fig.~\ref{fig:ens_weight}.
As can be seen, we dynamically assign weights for predicting samples of different neonates, and our promising ensemble performance indicating that in this way we can achieve better cross-person generalization.
Furthermore, our analysis reveals that \ours is assigned the highest weight for all neonates, further validating its usefulness in modeling neonatal physiological patterns.

\section{Conclusion}
This paper aims at neonatal seizure detection task based on EEG signals, which has been recognized as a common problem in NICU of the date while lack of enough attention.
We propose a spatial-temporal deep learning architecture which applies multi-channel spatial fusion with channel-level temporal modeling on EEG signal waves, tackling the challenges in practical scenario of neonatal brain health caring.
We also utilize an ensemble of diverse models to alleviate the generalization issue.
The experimental results on a large-scale real-world dataset of neonatal seizure detection have illustrated superior performance of our proposed solution with promising transferring ability on different EEG monitoring montages.

\bibliographystyle{IEEEtran}
\bibliography{main}

\end{document}